\documentclass{article}

\usepackage{arxiv}

\usepackage[utf8]{inputenc} 
\usepackage[T1]{fontenc}    
\usepackage{hyperref}       
\usepackage{url}            
\usepackage{booktabs}       
\usepackage{amsfonts}       
\usepackage{nicefrac}       
\usepackage{microtype}      
\usepackage{lipsum}
\usepackage{graphicx}
\usepackage{tanya_preamble}
\usepackage[group-separator={,},group-minimum-digits=4]{siunitx}
\usepackage[percent]{overpic}

\title{Accounting for Heavy Censoring in Evaluating the Risk Stratification Abilities of Existing Models for Time to Diagnosis of Huntington Disease}

\author{ 
\href{https://orcid.org/0000-0001-5927-3968}{\includegraphics[scale=0.06]{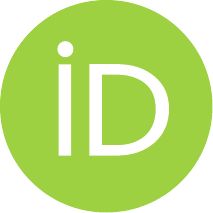}\hspace{1mm}Kyle F. ~Grosser$^{\dagger}$} \\
Department of Biostatistics, University of North Carolina at Chapel Hill, Chapel Hill, North Carolina, U.S.A.  \\
\And
\href{https://orcid.org/0000-0001-5510-8893}{\includegraphics[scale=0.06]{orcid.pdf}\hspace{1mm}Abigail G. ~Foes$^{\dagger}$} \\
Department of Biostatistics, University of North Carolina at Chapel Hill, Chapel Hill, North Carolina, U.S.A.  \\
\And
Stellen ~Li \\
Department of Biostatistics, University of North Carolina at Chapel Hill, Chapel Hill, North Carolina, U.S.A. \\
\And
Vraj ~Parikh \\
Department of Biostatistics, University of North Carolina at Chapel Hill, Chapel Hill, North Carolina, U.S.A. \\
\And 
\href{}{\includegraphics[scale=0.06]{orcid.pdf}\hspace{1mm}Tanya P.~Garcia} \\
Department of Biostatistics, University of North Carolina at Chapel Hill, Chapel Hill, North Carolina, U.S.A. \\
\And 
\href{https://orcid.org/0000-0001-5380-2427}{\includegraphics[scale=0.06]{orcid.pdf}\hspace{1mm}Sarah C.~Lotspeich} \\
Department of Statistical Sciences, Wake Forest University, Winston-Salem, North Carolina, U.S.A. \\
\texttt{lotspes@wfu.edu} \\
$^{\dagger}$These authors contributed equally to this work. \\
}

\begin{document}
\maketitle

\begin{abstract}
    Huntington disease (HD) is a neurodegenerative disease with progressively worsening symptoms. Accurately modeling time to HD diagnosis is essential for clinical trial design. Langbehn’s model, the CAG-Age Product (CAP) model, the Prognostic Index Normed (PIN) model, and the Multivariate Risk Score (MRS) model have all been proposed for this task. However, these models may yield conflicting predictions and few studies have systematically compared their performance. Further, those that have could be misleading due to (i) testing the models on the same data used to train them and (ii) failing to account for high rates of right censoring ($80\%+$) in performance metrics. 
    We discuss the theoretical foundations of these models, offering intuitive comparisons about their practical feasibility. We externally validate their risk stratification abilities using data from the ENROLL-HD study and two censoring-appropriate performance metrics, guiding the selection of a model for HD clinical trial design. As these models were originally developed in HD studies that ended more than a decade ago, we compared their predictive performance using published parameters versus updated ones (re-estimated using ENROLL-HD). We also show how these models can be used to estimate sample sizes for an HD clinical trial. 
    Based on either metric and using published or updated parameters, the MRS model, which incorporates the most covariates, performed best. 
    However, the much simpler PIN model offered similarly good performance while requiring fewer variables, many of which would require patients to undergo additional tests. In illustrating an HD clinical trial design, we defined an optimal threshold based on model performance metrics to determine which patients are more likely to be diagnosed. Sample size calculations using an optimal threshold based on metrics that did not account for censoring, as in previous studies, are shown to lead to underpowered trials. 
\end{abstract}


\section{Introduction}
\label{s:intro}

\subsection{Huntington Disease, Genetic Testing, and Observational Studies}
\label{sec:intro-aboutHD}

Huntington disease (HD) is a hereditary neurodegenerative disease caused by a repeat expansion of the trinucleotide CAG (cytosine-adenine-guanine) in the Huntingtin gene \citep{MacDonald1993}. This disease is fully penetrant for patients with $\geq 40$ CAG repeats, meaning that they will meet the diagnostic criteria for Huntington disease with probability $>90\%$ \citep{Tabrizi2022b}. An HD diagnosis is made based on a clinician's assessment of a patient's motor impairment, following the Unified Huntington Disease Rating Scale (UHDRS) \citep{Kremer1996}.

The full penetrance of HD is unique among neurodegenerative diseases. With genetic testing, researchers can identify patients who will develop HD before symptom onset and construct valuable observational cohorts (see Table~\ref{tab:HDstudies} for examples). These studies all collect crucial clinical measurements of HD symptom severity. The UHDRS Diagnostic Confidence Level (DCL) is a categorical rating from $0-4$ assigned by a clinician, where a DCL of $0$ indicates no motor impairment and $4$ indicates that the clinician is $99\%$ certain that the patient is exhibiting motor impairment due to HD \citep{Paulsen2019}. Crucially, a DCL rating of $4$ indicates a clinical HD diagnosis. The UHDRS Total Motor Score (TMS) \citep{Kremer1996} is a $31$-item, clinician-rated assessment of motor impairment, where higher scores indicate greater impairment. The Symbol Digit Modalities Test (SDMT) \citep{Smith1973} and Stroop Word, Color, and Interference tests \citep{Stroop1935} are timed assessments that quantify cognitive ability, with lower scores indicating worse cognitive impairment.

\begin{table}[ht]
\centering
\begin{tabular}{llll}
\hline
\textbf{Study} & \textbf{Years} & \textbf{Sample Size} & \textbf{Censoring Rate} \\
\hline
COHORT      & $2005$--$2011$     & \num{1326} & $81\%$ \\
TRACK-HD    & $2008$--$2011$     & 257        & $85\%$\\
PREDICT-HD  & $2002$--$2014$     & \num{1159} & $77\%$ \\
ENROLL-HD   & $2012$--Present  & \num{5173} & $88\%$ \\
\hline
\end{tabular}
\caption{Summary of major Huntington disease observational studies. 
COHORT \citep{Dorsey2012}; 
TRACK-HD \citep{Tabrizi2009}; 
PREDICT-HD \citep{Paulsen2006}; 
ENROLL-HD \citep{Sathe2021}.}
\label{tab:HDstudies}
\end{table}

\subsection{Designing Clinical Trials: Traditional Versus Preventative}

Clinically, HD is characterized by progressive motor dysfunction, cognitive decline, and loss of functional independence. Numerous pharmacological, psychiatric, and behavioral treatments have been proposed for HD patients, but none so far have been approved for delaying or stopping disease progression. However, several studies are in progress and one has shown promise in its human trials to reduce HD progression \citep{UCL2025_AMT130}. Another focus is on palliative care and improving quality of life, but treatments remain limited. Chorea, characterized by involuntary movements, is the only HD symptom with an approved pharmacological treatment \citep{OlmedoSaura2025}. Approved symptom treatments and recent clinical breakthroughs further highlight the need for early intervention. Risk stratification models are key to designing preventative clinical trials and identifying patients before symptoms manifest, allowing for potential interventions to have the most effect before irreversible symptoms and damage are incurred.

Traditionally, clinical trials for HD recruit patients who have already received their clinical diagnosis, and are thus displaying severe motor, cognitive, and functional impairment. This recruitment strategy allows for shorter trials and increases the likelihood of detecting treatment effects, which are both desirable traits to trial coordinators \citep{Langbehn2020}. However, this approach has a major limitation: By the time of diagnosis (and thus trial recruitment), neurological damage is irreversible and therapeutic interventions may be ineffective. Moreover, neurodegenerative diseases like HD progress slowly for decades prior to clinical diagnosis. Thus, trials that only recruit subjects after their clinical diagnosis could fail to detect interventions that impact the disease's early stages.

Given these limitations, researchers have begun to shift clinical trial focus to patients with prodromal HD, that is, those who have not yet been diagnosed but are genetically guaranteed to develop the disease \citep{Rubinsztein2016, Langbehn2020}. These studies, called \emph{preventative clinical trials}, aim to slow or stop symptom development and thus delay or prevent a patient's clinical HD diagnosis. 

However, one unique challenge with preventative trials is the small proportion of patients expected to be diagnosed during the trial. Paulsen et al. (2019) \cite{Paulsen2019} reported that less than $1\%$ of patients in an observational study of prodromal HD patients were diagnosed within $3$ years. If recruiting patients at all HD stages based on genetic testing, we would expect similar, low incidence of diagnosis in clinical trials, as they are often just a few years in length. Any enrolled patients who remain undiagnosed at trial end will have data that were administratively censored (i.e., ``cut off'') at their last follow-up visit. 

Since statistical power is tied to the number of ``events'' (HD diagnoses) in the untreated group, such trials often require large sample sizes to be sufficiently powered to detect a treatment effect. Put simply, when fewer patients are expected to be diagnosed during a trial, more must be enrolled to detect a treatment effect. With larger sample sizes, preventative trials can become more expensive and difficult to implement. 

Fortunately, time-to-diagnosis models can help to overcome this challenge via a strategy known as \emph{sample enrichment} \citep{Zhang2011, Long2017, Paulsen2019, Langbehn2022}. We can use these models to calculate log-risk scores for potential patients and recruit only those whose log-risk scores lie above a particular threshold (i.e., who are closer to diagnosis). This approach increases the expected rate of diagnosis in the untreated group, reducing the sample size needed for a desired statistical power. Sample enrichment makes preventative clinical trials for HD more cost-efficient and easier to implement.

\subsection{Comparing Existing Models for Time to Diagnosis}

Time-to-diagnosis models for HD are essential tools for both clinical trial planning and patient counseling. These models help estimate when an at-risk patient (based on genetic testing) will receive a clinical diagnosis. Currently, there exist four major time-to-diagnosis models for HD: Langbehn \citep{Langbehn2004}, CAG-Age Product (CAP) \citep{Zhang2011}, Multivariate Risk Score (MRS) \citep{Long2015}, and Prognostic Index Normed (PIN) \citep{Long2017}. All four are survival models built using data from observational studies of prodromal patients ($\geq 36$ CAG repeats), as described in Section~\ref{sec:ttd_models}. 

Having four competing models for predicting time to HD diagnosis raises an important question: Which one should we use when designing a preventative clinical trial? In theory, we want to use the ``best'' model, but there are several factors to consider, including model complexity, required input variables, and assumptions. This question would be best answered by a single analysis that directly compares the predictive performance of all four models using the same dataset. Specifically, they should be compared using a new dataset on which none of the four models was developed (i.e., externally validated). 

As these models were developed, some comparisons were made to the existing models, but none considered all four. The Langbehn model was developed first, so there were no comparisons to make \citep{Langbehn2004}. The CAP model came next, and Zhang et al. (2011) \cite{Zhang2011} discussed the Langbehn model but did not provide quantitative comparisons. Long et al. (2015) \cite{Long2015} compared the newly-proposed MRS model to a nonparametric model similar to CAP but again made no comparison to Langbehn. Finally, Long et al. (2017) \cite{Long2017} compared the PIN model to a semiparametric model similar to CAP, but made no comparison to Langbehn or MRS.

Despite the public availability of these modeling tools, limited work has evaluated and compared them systematically. One of the few comparisons comes from Paulsen et al. (2019) \cite{Paulsen2019}, who assessed the predictive performance of the CAP, MRS, and PIN models using receiver operating characteristic (ROC) analysis. This technique evaluates how well a model separates patients who will be diagnosed up to versus after a specific time. There are three key limitations to the model evaluation conducted by Paulsen et al. (2019) \cite{Paulsen2019} 

First, only three of the four major models were assessed. Second, their ROC analyses did not account for the high proportion of patients in their data who remained undiagnosed at study end (i.e., whose times to diagnosis were \emph{right censored}). Although the time-to-diagnosis models were designed to handle censored data, this validation approach was not. Standard ROC analyses assume that all patients' diagnosis times are known. Third, there was \emph{double-dipping} \citep{Ball2020}, because the models were evaluated on the PREDICT-HD dataset \citep{Paulsen2006} from which they were built. (The CAP and MRS models were developed from PREDICT-HD alone; the PIN model was developed on a combination of studies that included PREDICT-HD.) 

Additional work by Long et al. (2023) \cite{Long2023} examined the PIN and CAP models in the context of the Huntington Disease Integrated Staging System (HD-ISS), which can be used as an alternative outcome measure to time to clinical diagnosis in Huntington disease clinical trials \citep{Tabrizi2022b}. Long et al. (2023) \cite{Long2023} also used ROC analysis to compare model performance, incurring similar limitations to Paulsen et al. (2019) \cite{Paulsen2019} due to censoring. Moreover, their comparisons excluded the Langbehn and MRS models, and performance in predicting HD-ISS does not generalize to other HD outcomes.

Together, these limitations weaken the reliability and generalizability of previous comparisons, motivating the need for a comprehensive comparison of the four major time-to-diagnosis models for HD. In particular, censoring-appropriate metrics are needed for performance evaluation, and the data used must be independent of each model's development.

\subsection{Overview}

Multiple time-to-diagnosis models are available for HD, each relying on distinct assumptions and clinical inputs. To aid patients, clinicians, and researchers in deciding between these models, accurate and exhaustive comparisons of predictive performance are needed. We apply the four most common time-to-diagnosis models for HD (Langbehn, CAP, MRS, and PIN) to a single dataset that is independent of each model's development (the ENROLL-HD study). Using censoring-appropriate metrics, Uno's concordance statistic (C statistic) and Kaplan-Meier-based ROC analysis, we compare their predictive performance. Our comparisons provide guidance to researchers in determining which model is best for their situation. Finally, we demonstrate how the best-performing model can be used in sample enrichment for a preventative clinical trial. The rest of the paper is organized as follows. In Section~\ref{methods}, the models are discussed in more detail and methods for evaluation are introduced. The application to ENROLL-HD, including sample size calculations, is presented in Section~\ref{results}. We conclude with a brief discussion and model selection recommendations in Section~\ref{discuss}.

\section{Methods}\label{methods}

\subsection{Notation}
\label{subsec:notation}
Suppose we observe data for $n$ patients, indexed by $i=1,...,n$. Let $X_i$ denote patient $i$'s continuous diagnosis time and $\bZ_i$ denote their covariate vector. In survival analysis, our goal is typically to estimate some aspect of the conditional distribution of $X_i$ given $\bZ_i$. When the primary goal is sample enrichment, the \emph{log-risk score} for $X_i$ given $\bZ_i$ is the estimand of interest, denoted by $\bbeta\trans \bZ_i$ for some model-specific parameter vector $\bbeta$ that describes this conditional distribution. This score is a scalar summary of patient $i$'s risk of being diagnosed with HD, where higher scores indicate greater risk \citep{Park2021}. 

However, analysts often encounter data in which $X_i$ is censored by another variable $C_i$ for some patients. We focus on random right censoring, where $X_i$ is censored by the random variable $C_i$ if $X_i > C_i$. In many HD studies, more than $75\%$ of patients' diagnosis times are randomly right censored (Table~\ref{tab:HDstudies}). Typically, censored patients dropped out of a study before being diagnosed or the study ended before they were diagnosed, in which case $C_i$ would be time to drop out or study end, respectively. Patient deaths during the trial are possible but handled in the same way as drop-outs or those censored at study end for analyses. Because preventative clinical trials focus on patients who have not yet received an HD diagnosis, any deaths observed during the trial would likely be attributed to other, non-HD causes, avoiding concerns about competing risks. Hence, we have $W_i=\min(X_i,C_i)$ and $\Delta_i=\textrm{I}(X_i\leq C_i)$ so that our observed, patient-level data are $(W_i, \Delta_i, \bZ_i)$. Our goal, then, is to estimate log-risk scores $\bbeta\trans \bZ_i$ using these data, though our true interest is in the full data pair $(X_i, \bZ_i)$.

\subsection{Time-to-Diagnosis Models}
\label{sec:ttd_models}

\subsubsection{Model 1: Langbehn}

The Langbehn model was introduced to estimate the probability of HD diagnosis at a given age as a function of CAG repeats \citep{Langbehn2004}. It was developed using retrospective data collected from $40$ clinical sites across Africa, Asia, Europe, and North America. While this dataset did not come from one of the commonly cited cohorts in Table~\ref{tab:HDstudies}, it constituted the largest dataset of HD patients to date.

When developing this model, Langbehn et al. (2004) \cite{Langbehn2004} restricted the dataset to patients with $41$ to $56$ CAG repeats (inclusive). After these exclusions, they had $2913$ patients, of whom $2298$ had been clinically diagnosed by study end. With 615 patients who had not been diagnosed (i.e., whose times of diagnosis had been right-censored), the sample saw a right-censoring rate of $21\%$. This proportion contrasts sharply with the cohorts described in Table~\ref{tab:HDstudies}, which saw between $77\%$ and $88\%$ undiagnosed patients.

To develop the Langbehn model, a parametric survival modeling framework was used. Specifically, this model assumes that age at HD diagnosis ($X_i$) follows a logistic distribution conditional on CAG repeats ($\text{CAG}_i$). Following from the assumed distribution, the \emph{Langbehn model} expresses the conditional probability that a patient is diagnosed by age $x$ given a particular CAG repeats $\textrm{CAG}_i$ as
\begin{align}
\label{eqn:Langbehn_model}
    {\Pr}_{\bbeta,\bgamma}(X_i\leq x\mid \textrm{CAG}_i) = \left[1 + \exp\left\{ -\left(\frac{\pi}{\sqrt{3}}\right) \frac{x - \mu(\textrm{CAG}_i;\bbeta)}{\sqrt{\sigma^2(\textrm{CAG}_i;\bgamma)}} \right\}\right]^{-1}, 
\end{align}
where $\mu(\textrm{CAG}_i;\bbeta) = \beta_0 + \exp(\beta_1 - \beta_2 \textrm{CAG}_i)$ and $\sigma^2(\textrm{CAG}_i;\bgamma) = {\gamma_0 + \exp(\gamma_1 - \gamma_2 \textrm{CAG}_i)}$ are the mean age at diagnosis and its variance, respectively (both dependent on $\textrm{CAG}_i$). The parameters ($\bbeta$, $\bgamma$) were estimated via maximum likelihood estimation (MLE), accounting for the right-censoring in age at HD diagnosis. In addition to assuming a logistic distribution, the Langbehn model assumes that $X_i$ is conditionally independent of $C_i$ given the patient's CAG repeats (i.e., noninformative censoring).

The Langbehn model can also provide time-to-diagnosis predictions. Building upon \eqref{eqn:Langbehn_model}, Langbehn et al. (2004) \cite{Langbehn2004} calculated the conditional probability that patient $i$ will be diagnosed by a future $x_f$, given their CAG repeats $\textrm{CAG}_i$ and current age $x_c$, as
\begin{align}
{\Pr}_{\bbeta,\bgamma}(X\leq x_f \mid \textrm{CAG}_i, x_c) &= 
1-\frac{1-{\Pr}_{\bbeta,\bgamma}(X \leq x_f \mid \textrm{CAG}_i)}{1-{\Pr}_{\bbeta,\bgamma}(X\leq x_c \mid \textrm{CAG}_i)}, \label{eqn:Langbehn_model_condl}
\end{align}
assuming they have not yet been diagnosed at age $x_c$.

Crucially, the Langbehn model does not produce a log-risk score like the other models do. However, \eqref{eqn:Langbehn_model_condl} provides a single-number summary of a patient's short-term risk of diagnosis given their age and genetic information and can be used to stratify patients by risk, which aligns the Langbehn model’s output with the other models and facilitating direct comparison. The Langbehn model cannot be estimated using existing software; we have developed R code to accompany this paper \citep{R-software}.

\subsubsection{Model 2: CAG-Age Product (CAP)}

The CAP model was introduced by Zhang et al. (2011) \cite{Zhang2011} to predict a patient's time to diagnosis given their CAG repeats and current age. It was developed using data from the PREDICT-HD study, restricted to patients with (i) a CAG repeats $\geq 36$ and (ii) DCL $<4$ at study entry. That is, patients needed to be at-risk for Huntington disease but not yet diagnosed at enrollment. After applying these criteria, there were $730$ patients. During the study, $137$ of these patients were clinically diagnosed, while the remaining $593$ ($81\%$) were not and thus had right-censored times of diagnosis.

The CAP model is a parametric accelerated failure time model with time to diagnosis ($X_i$) as the outcome and current age and CAG repeats as covariates. During development, candidate models using all combinations of covariates and their interaction were evaluated, in conjunction with five  distributional assumptions for the error term. Candidate models were evaluated using Akaike's information criteria and prediction error. The best-fitting model had (i) current age and its interaction with CAG repeats as covariates and (ii) assumed a logistic distribution for the error (admitting a log-logistic distribution for $X_i$ given covariates).

The \emph{CAP model} is defined $\log(X_i) = \beta_0 + \beta_1 \text{Age}_i(\text{CAG}_i+\beta_2) + \epsilon,$ where $\bbeta = (\beta_0, \beta_1, \beta_2)$ denote the intercept and effects of $\text{Age}_i \times \text{CAG}_i$ and $\text{Age}_i$. The error term $\epsilon$ is assumed to follow a logistic distribution with location $=0$ and scale $=\sigma$. The linear predictor, $\beta_0 + \beta_1 \text{Age}_i(\text{CAG}_i+\beta_2)$, can be interpreted as a log-risk score. The parameters ($\bbeta$, $\sigma$) are estimated via MLE, with a likelihood function that accounts for right censoring. Time to diagnosis $X_i$ is assumed to be conditionally independent of $C_i$ given CAG repeats and current age. The CAP model can be estimated using the \texttt{survreg()} function from the \texttt{survival} package in R with option \texttt{dist = "loglogistic"} \citep{Therneau2022}.

\subsubsection{Model 3: Multivariate Risk Score (MRS)}

The MRS model was developed with similar methods and intended applications to the CAP model. Designed with observational studies in mind, the MRS model predicts the time from enrollment to HD diagnosis \citep{Long2015}. However, compared to Langbehn and CAP, MRS was developed using a much broader set of motor, imaging, cognitive, functional, psychiatric, and demographic covariates.

The MRS model was developed using data from PREDICT-HD, subsetting to patients with (i) CAG repeats $\geq 36$ and (ii) DCL $<4$ at baseline. With these criteria (the same as those for the CAP model), the analytic dataset had a sample size of $1078$ with a censoring rate of $79\%$, since $225$ patients were clinically diagnosed during the study while $853$ were not.

Unlike the models discussed thus far, the MRS model is semiparametric. Semiparametric models avoid assuming a specific distribution for time to diagnosis $X_i$ (e.g., log-logistic). Also, a much larger set of covariates $\bZ_i$ was considered. Common measures like the TMS and DCL were included, as were brain region measurements, motor assessments, biological sex, and $25$ other variables. 

To develop the MRS model, Long et al. (2015) \cite{Long2015} began by conducting a variable selection procedure. They used random survival forests, a nonparametric technique that accommodates non-linear and interaction effects while avoiding specific distributional assumptions about $X_i$. Candidate models with varying sets of covariates were evaluated based on an inverse probability of censoring (IPCW)-based Brier score \citep{Park2021}. 

After selecting covariates, a Cox proportional hazards model was fit using the most important main effects and interactions. The \emph{MRS model} was specified as $\lambda_X(t \mid \bZ_i) =$
\begin{align*}
    \lambda_0(t) \exp(
    &\beta_1 \text{DCL}1_i + \beta_2 \text{DCL}2_i + \beta_3 \text{DCL}3_i + \beta_4 \text{TMS}_i + \beta_5 \text{Color}_i + \\ 
    & \beta_6 \text{Word}_i + \beta_7 \text{Interference}_i + \beta_8 \text{SDMT}_i + \beta_9 \text{CAG}_i + \\
    & \beta_{10} \text{Age}_i + \beta_{11} \text{TMS}_i^2 + \beta_{12} \text{CAG}_i \times \text{TMS}_i + \beta_{13} \text{CAG}_i \times \text{Age}_i)
\end{align*}
where $\lambda_X(t \mid \bZ)$ denotes the conditional hazard function of $X$ given $\bZ$ evaluated at $t$; $\lambda_0(t)$ is the baseline hazard function, the hazard function of $X$ when the covariates are all zero ($\bZ_i=\bzero$); and $\text{DCL}1$, $\text{DCL}2$, and $\text{DCL}3$ are indicators of DCLs $= 1$, $2$, and $3$, respectively. The coefficients $\bbeta = (\beta_1,\ldots,\beta_{13})\trans$ represent adjusted log hazard ratios for the covariates and were estimated via semiparametric MLE, since $\lambda_0(t)$ is estimated nonparametrically.

Like Langbehn and CAP, the MRS model assumes that $X$ is conditionally independent of $C$ given $\bZ$. It also uniquely assumes that the hazard ratio between two patients with different covariates remains constant over time (i.e., proportional hazards). Similar to the CAP model, the linear predictor of the MRS model, $\bbeta\trans \bZ$, is interpretable as a log-risk score, which enables risk stratification. Moreover, the model can be estimated using the \texttt{coxph()} function from the \texttt{survival} package in R \citep{Therneau2022}.

\subsubsection{Model 4: Prognostic Index Normed (PIN)}
\label{sec:methods_PIN}

Most recently, Long et al. (2017) \cite{Long2017} introduced the PIN model, which is another Cox model of time to HD diagnosis. In contrast to the Langbehn, CAP, and MRS models, the PIN model was developed using integrated data from three studies: PREDICT-HD, COHORT, and TRACK-HD. Each dataset was restricted to include only patients who met the following criteria at enrollment: (i) $\text{CAG}\geq 36$ and (ii) $\text{DCL}<4$. These criteria, which are identical to those for CAP and MRS, left a total sample size of $1421$ patients, with a right-censoring rate of $77\%$.

Development of the PIN model began with a variable selection procedure. Long et al. (2017) \cite{Long2017} fit ten pre-planned Cox regression models (drawing from their earlier HD research) with time to diagnosis $X_i$ as the outcome and different combinations of seven covariates: DCL, TMS, Stroop, SDMT, CAG, Age, and $\textrm{CAP}=\textrm{Age}(\textrm{CAG}-34)$. These candidate models were fit using only the PREDICT-HD data, and Harrell's C statistic was used to evaluate each model's performance. To choose from among the ten covariate combinations, leave-one-site-out cross validation was used. That is, the candidate models were estimated using data from all clinical sites in PREDICT-HD, omitting one, and then its predictive performance was measured via Harrell's C using data from the held-out site. 

The model with the highest median Harrell's C statistic across sites was selected, which consisted of main effects for TMS, SDMT, and CAP (no interactions). The final PIN model was specified as $\lambda_X(t \mid \bZ_i) = \lambda_0(t) \exp(\beta_1 \textrm{TMS}_i + \beta_2 \textrm{SDMT}_i + \beta_3 \textrm{CAP}_i)$, where $\lambda_X(t \mid \bZ)$, $\lambda_0(t)$, and $\bbeta = (\beta_1,\beta_2,\beta_3)\trans$ have the same interpretations as in the MRS model. The log hazard ratios $\bbeta$ were estimated using combined data from the PREDICT-HD, COHORT, and TRACK-HD studies.

Since PIN is also a Cox model, it shares several properties with other common models. First, it makes similar assumptions: (i) $X$ is conditionally independent of $C$ given $\bZ$ and (ii) proportional hazards. Second, $\bbeta\trans \bZ$ from the PIN model is interpretable as a log-risk score and can be used to stratify patients by risk. The model can be estimated using the \texttt{coxph()} function from the \texttt{survival} package in R \citep{Therneau2022}.

\subsection{Modeling Trade-Offs}
\begin{table}[ht]
    \centering
    \label{tab:summary_of_models}
    \begin{tabular}{
        >{\raggedright\arraybackslash}p{2cm}
        >{\raggedright\arraybackslash}p{2.75cm}
        >{\raggedright\arraybackslash}p{2.75cm}
        >{\raggedright\arraybackslash}p{3cm}
    }
        \toprule
        \textbf{Model} & \textbf{Covariates (number)} & \textbf{Type of model} & \textbf{Dataset(s) used} \\
        \midrule
        Langbehn & CAG repeat length, age (2) & Parametric; not an ``off-the-shelf'' method& Retrospective clinical data \\
        \addlinespace
        CAP & Same as Langbehn (2) & Parametric; accelerated failure time& PREDICT-HD \\
        \addlinespace
        PIN & Same as CAP plus TMS and SDMT (4)& Semiparametric; Cox proportional hazards& PREDICT-HD, COHORT, and TRACK-HD \\
        \addlinespace
        MRS & Same as PIN plus DCL and Stroop test scores (8)& Semiparametric; Cox proportional hazards& PREDICT-HD \\
        \bottomrule
    \end{tabular}
    \caption{Summary of existing models for time to diagnosis of Huntington disease: Langbehn, CAG-Age Product (CAP), Multivariate Risk Score (MRS), and Prognostic Index Normed (PIN). See \textbf{Figures~\ref{fig:stackUnoC}} and \textbf{\ref{fig:stackAUROC}} for time-specific evaluation of model performance in the ENROLL-HD study.}
\end{table}
Recall that our ultimate goal with these models is to stratify patients by risk, facilitating sample enrichment for the design of preventative clinical trials. In this section, we discuss two qualitative considerations that arise when deciding on a time-to-diagnosis model: (i) the covariates required and (ii) the statistical assumptions imposed. An empirical comparison of these models when they are applied to data from the ENROLL-HD study follows in Section~\ref{results}. 


Both the Langbehn and CAP models incorporate a patient's age at enrollment and CAG repeats. This small number of required variables makes these models simple to implement. However, these models omit other potentially informative covariates, like motor and cognitive test scores. In contrast, both the MRS and PIN models include the TMS and SDMT score, and the MRS model further includes DCL and the Stroop test scores. 

Including more covariates in time-to-diagnosis models could increase the costs of data collection and complicate implementation, but it could also enable more nuanced risk stratification. For example, consider two patients whose CAG repeats and age at enrollment are identical. These patients would have the same estimated log-risk score from either the Lanbehn or CAP models, since they rely on just these two covariates. (The scores from these two models will not necessarily be the same for these patients.) However, if these patients had distinct SDMT scores, the PIN and MRS models would produce different log-risk scores for them, reflecting the added information and allowing us to prioritize them separately. 


We also consider the statistical assumptions made by each model. The Langbehn and CAP models are both fully parametric. The Langbehn assumes that $X$ follows a logistic distribution, and the CAP model assumes that it follows a log-logistic distribution. If either model is used to make predictions about data that do not satisfy these assumptions, then the resulting parameter estimates could suffer from bias, leading to poor predictions and misleading risk stratification. Although the MRS and PIN models do not impose distributional assumptions on $X$, they both rely on the proportional hazards assumption. Violation of this assumption could have the same repercussions. However, assuming proportional hazards may be less restrictive than fully specifying a distribution.

\subsection{External Validation Versus Model Updating}

In practice, it would be simplest to apply a published model ``as-is'' to new data, and treat the predictions as established equations. For example, we could predict with $\log(\widehat{X}_i) = \hat{\beta}_0 + \hat{\beta}_1 \textrm{Age}_i(\textrm{CAG}_i+\hat{\beta}_2)$ following the CAP model where the estimates $\hat{\beta}_0,\hat{\beta}_1,\hat{\beta}_2$ were estimated using the PREDICT-HD study and can be found in Zhange et al. (2011) \cite{Zhang2011} However, since the models thus far have only been evaluated on the same data used to train them (i.e., subject to double-dipping), it is unclear how well they would perform in a new cohort of HD patients. \textit{External validation}, wherein the published coefficient estimates (based on the original training data) are applied to new testing data (ENROLL-HD), is needed to evaluate the transportability of the four popular HD models \citep{Moons2012}. 

Moreover, the models were all trained on HD cohorts that ended more than $10$ years ago (Table~\ref{tab:HDstudies}), whereas many more recent studies, like ENROLL-HD, are still ongoing. Therefore, we also considered \textit{model updating}, wherein the predictors and distributions (where applicable) of all models were unchanged but the corresponding coefficients were re-estimated (updated) using the ENROLL-HD data. We used $5$-fold cross-validation to evaluate the performance of the updated models, pooling the metrics from Section~\ref{sec:metrics} across folds by taking the means. These pooled metrics should be more representative of performance in other HD studies than those obtained from a single fit, since the latter would introduce double-dipping by training and testing on ENROLL-HD. 

\subsection{Metrics for Evaluating Risk Stratification}
\label{sec:metrics}

We considered two censoring-adjusted metrics to evaluate the models' risk stratification ability. First, Uno's Concordance (Uno's C) statistic assesses how well the models rank pairs of patients in terms of log-risk scores. Second, Kaplan-Meier-adjusted receiver operating characteristic (ROC) curves and their corresponding areas under the curve (AUC) assess how well the models separate the group of patients diagnosed by a particular time $\tau$ from the group undiagnosed by time $\tau$.

\subsubsection{Metric 1: Uno's Concordance (Uno's C) Statistic}

To measure predictive model performance with heavily censored data, Uno et al. (2011) \cite{Uno2011} introduced a novel concordance index that incorporates IPCW to make use of the full dataset. Specifically, \emph{Uno's C statistic} weights each pair ($i$, $j$) of uncensored observations ($i = 1, \dots, n$; $j = 1, \dots, n$; $i\neq j$) based on the estimated probability that the patient with the smaller observed time $W$ (time to diagnosis or censoring) remains uncensored up to $\min(W_i,W_j)$. These weights are calculated from the Kaplan-Meier survival estimator for $C$, denoted by $\widehat{G}(c) = \widehat{\Pr}(C>c)$. 

The central quantity in Uno's C statistic is $\bbeta\trans\bZ_i$, which denotes patient $i$'s log-risk score given covariates $\bZ_i$ and parameters $\bbeta$. The Langbehn model does not produce a log-risk score, but we can use the conditional probability from \eqref{eqn:Langbehn_model_condl} as a proxy when calculating its Uno's C statistic. Letting $\hat{\bbeta}\trans\bZ_i$ denote the estimated log-risk score for patient $i$ given parameter estimates $\hat{\bbeta}$ from a particular model, Uno's C statistic is calculated as
\begin{align*}
C_{\text{U}}(\hat{\bbeta}, \tau) = \frac{
\sum_{i=1}^{n} \sum_{j\neq i} \Delta_i \left\{ \widehat{G}(W_i)^{-2}\right\}
\text{I}\left(W_i < W_j, W_i < \tau\right) {\rm I}\left(\hat{\bbeta}\trans\bZ_i>\hat{\bbeta}\trans\bZ_j\right)  
}{
\sum_{i=1}^{n} \sum_{j\neq i} \Delta_i \left\{ \widehat{G}(W_i)^{-2} \right\}
\text{I}\left(W_i < W_j, W_i < \tau\right)
},
\end{align*}
where $\tau$ is the time horizon of interest such that $\widehat{G}(\tau)>0$. We calculate $C_{\text{U}}(\hat{\bbeta}, \tau)$ for a range of $\tau$ values to compare the models across several time horizons. Available software includes the \texttt{UnoC()} function from the \texttt{survAUC} package in R \citep{Potapov2024}.

Formally, Uno's C statistic estimates the probability that a randomly selected patient who was diagnosed before time $\tau$ and earlier than another patient is assigned a higher log-risk score than that other patient. Uno's C estimates close to $1$ indicate strong risk stratification, whereas values below $0.5$ indicate model performance that is worse than a coin flip. Uno's C statistic makes greater use of the available data by incorporating $\widehat{G}(c)$ as weights, making it more robust to heavy censoring than Harrell's C. Although Uno's C still discards some ``incompatible pairs'' (i.e., any $[i, j]$ with both patients censored, or $W_i<W_j$ but $\Delta_i=0$), these pairs still indirectly contribute to Uno's C statistic by informing $\widehat{G}(c)$.

\subsubsection{Metric 2: Receiver Operating Characteristic (ROC) Curves}

In survival analysis, time-dependent ROC curves are also used to evaluate a model’s risk stratification ability at specific time points. These curves plot a model's sensitivity (true positive rate) on the y-axis against $(1-\text{specificity})$ (false positive rate) on the x-axis at a given time $t$, indicating how well it distinguishes between patients who are diagnosed by time $t$ and those who are not. The true positive rate ($\text{TPR}(q,t)$) and false positive rate ($\text{FPR}(q,t)$) for threshold $q$ at time $t$ are defined as $\text{TPR}(q,t) = \Pr(\bbetahat\trans \bZ_i > q \mid X_i \leq t)$  and $\text{FPR}(q,t) = \Pr(\bbetahat\trans \bZ_i > q \mid X_i > t)$, where $\bbetahat\trans \bZ_i$ continues to denote the estimated log-risk score for patient $i$ from a given model. These quantities represent the proportions of patients who are correctly or incorrectly classified as high risk (i.e., $\bbetahat\trans \bZ_i > q$) at time $t$, based on whether they have been diagnosed by that time (i.e., $X_i\leq t$). 

In practice, we calculate $\text{TPR}(q,t)$ and $\text{FPR}(q,t)$ at thresholds $q$ ranging from $\min(\bbetahat\trans\bZ_1,\ldots,\bbetahat\trans\bZ_n)$ to $ \max(\bbetahat\trans\bZ_1,\ldots,\bbetahat\trans\bZ_n)$. We obtain the ROC curve at time $t$ by plotting the $\{\text{FPR}(q,t), \text{TPR}(q,t)\}$ pairs over the range of thresholds $q$. By evaluating ROC curves at multiple time points (e.g., at $3$, $5$, or $10$ years), we can assess a model's performance across the follow-up period.

In the absence of censoring, we could estimate $\text{TPR}(q,t)$ and $\text{FPR}(q,t)$ using sample proportions as
\begin{align*}
    \widetilde{\text{TPR}}(q, t) &= \sum_{i=1}^n \text{I}\left(X_i \le t, \bbetahat\trans\bZ_i > q\right)/\sum_{i=1}^n \text{I}(X_i \le t), \;\text{and}\\
    \widetilde{\text{FPR}}(q, t) &= \sum_{i=1}^n \text{I}\left(X_i > t, \bbetahat\trans\bZ_i > q\right)/\sum_{i=1}^n \text{I}(X_i > t).
\end{align*}
When censoring is present, the same formulas could be applied to only the subset of uncensored observations, essentially leading to complete-case estimates of the TPR and FPR. However, such metrics could only describe the model's risk stratification ability among diagnosed patients. Particularly in observational data of HD, when $X_i$ is heavily right-censored, $\widetilde{\text{TPR}}(q, t)$ and $\widetilde{\text{FPR}}(q, t)$ would not comprehensively describe model performance across the entire sample. 

Heagerty et al. (2000) \cite{heagerty2000time} proposed censoring-adjusted estimators 
that incorporate censored observations and describe model performance for all patients: 
\begin{align*}
\widehat{\text{TPR}}(q, t) & = \frac{\left\{ 1 - \widehat{S}\left(t \mid \bbetahat\trans \bZ_i > q\right) \right\} \widehat{\Pr}\left(\bbetahat\trans \bZ_i > q\right)}{1 - \widehat{S}(t)} \text{ and } \\
\widehat{\text{FPR}}(q, t) & = 1 - \frac{\widehat{S}\left(t \mid \bbetahat\trans \bZ_i \leq q\right) \widehat{\Pr}\left(\bbetahat\trans \bZ_i \leq q\right)}{\widehat{S}(t)},
\end{align*}
where $\widehat{S}(t \mid \bbetahat\trans \bZ_i > q)$ and $\widehat{S}(t \mid \bbetahat\trans \bZ_i \leq q)$ denote the conditional Kaplan-Meier survival estimators for $X$ among patients whose log-risk scores exceed or fall below $q$, respectively; $\widehat{\Pr}(\bbeta\trans \bZ_i > q)$ and $\widehat{\Pr}(\bbeta\trans \bZ_i \leq q)$ denote the sample proportions in these two groups; and $\widehat{S}(t)$ denotes the marginal Kaplan-Meier survival estimator for $X$ evaluated at $t$.

To summarize a model's risk stratification ability based on ROC analysis, we can calculate the AUC. AUC estimates take on values in the range $[0,1]$, with larger values indicating stronger risk stratification and values $< 0.5$ indicating performance worse than a coin flip.  To compare the four models, we use the AUC of the Kaplan-Meier-adjusted ROC curves, specifically. 

\section{Results}
\label{results}

\subsection{Data Description}
To compare the four time-to-diagnosis models, we turned to data from the ENROLL-HD study, which were not used to develop any of the models. ENROLL-HD is a longitudinal observational study developed to improve scientific understanding of HD. Patients from three risk groups were enrolled in this study: (i) those who were clinically diagnosed with HD, (ii) those with CAG repeats $\geq36$ but who were not yet diagnosed, and (iii) those with CAG repeats $\leq36$ \citep{Sathe2021}. The study is massive in scope, with $\num{21818}$ current patients enrolled across $156$ clinical sites (mostly in Europe and North America, with some in Latin America and Oceania). Given its broad patient population and global scope, ENROLL-HD serves as a powerful research platform for studying HD.

The ENROLL-HD data include demographic and clinical information on all variables required to implement the four models under consideration (Langbehn, CAP, MRS, and PIN). Specifically, we extracted age at enrollment, CAG repeats, UHDRS DCL, UHDRS TMS, SDMT score, and Stroop test scores. Before comparing the four models, we filtered the ENROLL-HD data to include only (i) patients with $40$--$57$ CAG repeats (ii) who were not yet clinically diagnosed with HD at their first visit. Filtering the data in this way led to an analytic sample of $n=3113$ patients, of whom $725$ were clinically diagnosed during the study, yielding a censoring rate of $77\%$. These patients are summarized in Table~\ref{tab:table1_strat}. 
\begin{table}[ht]
\centering
\begin{tabular}[t]{lll}
\toprule
\textbf{Variable} & \textbf{Undiagnosed} & \textbf{Diagnosed} \\
& \textbf{(Right-censored)} & \textbf{(Uncensored)} \\
\midrule
Number of patients & 2388 & 725\\
Age at enrollment & 36.25 (9.94) & 44.07 (11.48)\\
Sex & & \\
\hspace*{1.5em} Male & 951 (39.8\%) & 334 (46.1\%)\\
\hspace*{1.5em} Female & 1437 (60.2\%) & 391 (53.9\%)\\
CAG repeats & 43.23 (2.23) & 44.02 (2.90)\\
UHDRS DCL & & \\ 
\hspace*{1.5em} 0 & 1515 (63.4\%) & 121 (16.7\%)\\
\hspace*{1.5em} 1 & 581 (24.3\%) & 191 (26.3\%)\\
\hspace*{1.5em} 2 & 184 ( 7.7\%) & 187 (25.8\%)\\
\hspace*{1.5em} 3 & 108 ( 4.5\%) & 226 (31.2\%)\\
UHDRS TMS & 2.52 (3.87) & 9.56 (9.35)\\
SDMT score & 50.77 (11.62) & 39.15 (12.58)\\
Stroop Tests & & \\
\hspace*{1.5em} Word Reading score & 94.36 (17.72) & 79.47 (20.23)\\
\hspace*{1.5em} Color Naming score & 73.48 (14.62) & 62.10 (16.20)\\
\hspace*{1.5em} Interference score & 44.50 (11.16) & 35.96 (11.86)\\
\bottomrule
\end{tabular}
\caption[Summary of Demographic and Clinical Variables]{Baseline characteristics of patients in the ENROLL-HD analytic sample, stratified by diagnostic status. Means (standard deviations) are shown for continuous variables; counts (percentages) are shown for categorical variables. Age is in years. CAG: Cytosine-adenine-guanine. UHDRS: Unified Huntington Disease Rating Scale. DCL: Diagnostic Confidence Level. TMS: Total Motor Score. SDMT: Symbol Digit Modalities Test.}
\label{tab:table1_strat}
\end{table}
\subsection{Comparing Risk Stratification Abilities}\label{sec:compare_risk_strat}

We present two comparative analyses of the four models. First, we applied each model with the original published parameters (external validation). Second, we evaluated each using the published distributions (if applicable) and covariates but updated parameters based on ENROLL-HD (model updating). The first comparison gauged the models' transportability to new HD studies, and the second assessed whether tailoring coefficients to new data but retaining the original structure could further improve performance. In each of the following subsections, we begin by discussing cross-validation with the updated models and end by comparing them to the originals. Ultimately, we found that all models performed similarly with original or updated coefficients. 

\subsubsection{Uno's C Statistics}

To evaluate the performance with updated parameters tailored to the ENROLL-HD data, we used a $5$-fold cross-validation procedure. We randomly split the data into $5$ mutually exclusive subgroups (``folds''). Selection of the folds was stratified so that each fold contained $725/5 = 145$ uncensored patients, preserving the sample's censoring rate ($78\%$) within each.

Iterating through the $k$ folds ($k = 1, \dots, 5$), we estimated (trained) each model's parameters $\bbetahat$ using data from all folds except $k$ and then evaluated (tested) those parameters' performance on the $k$th fold. When evaluating each split, we calculated time-specific Uno's C statistics for all follow-up times in addition to a global statistic. The global statistic is a weighted average of the time-specific statistics, with weights given by the empirical mass function of the event times at each evaluation time $\tau$. With cross-validation, we approximated the models' performance when applied to data external to the training set.

\begin{figure}
    \centering
    \includegraphics[width=0.8\linewidth]{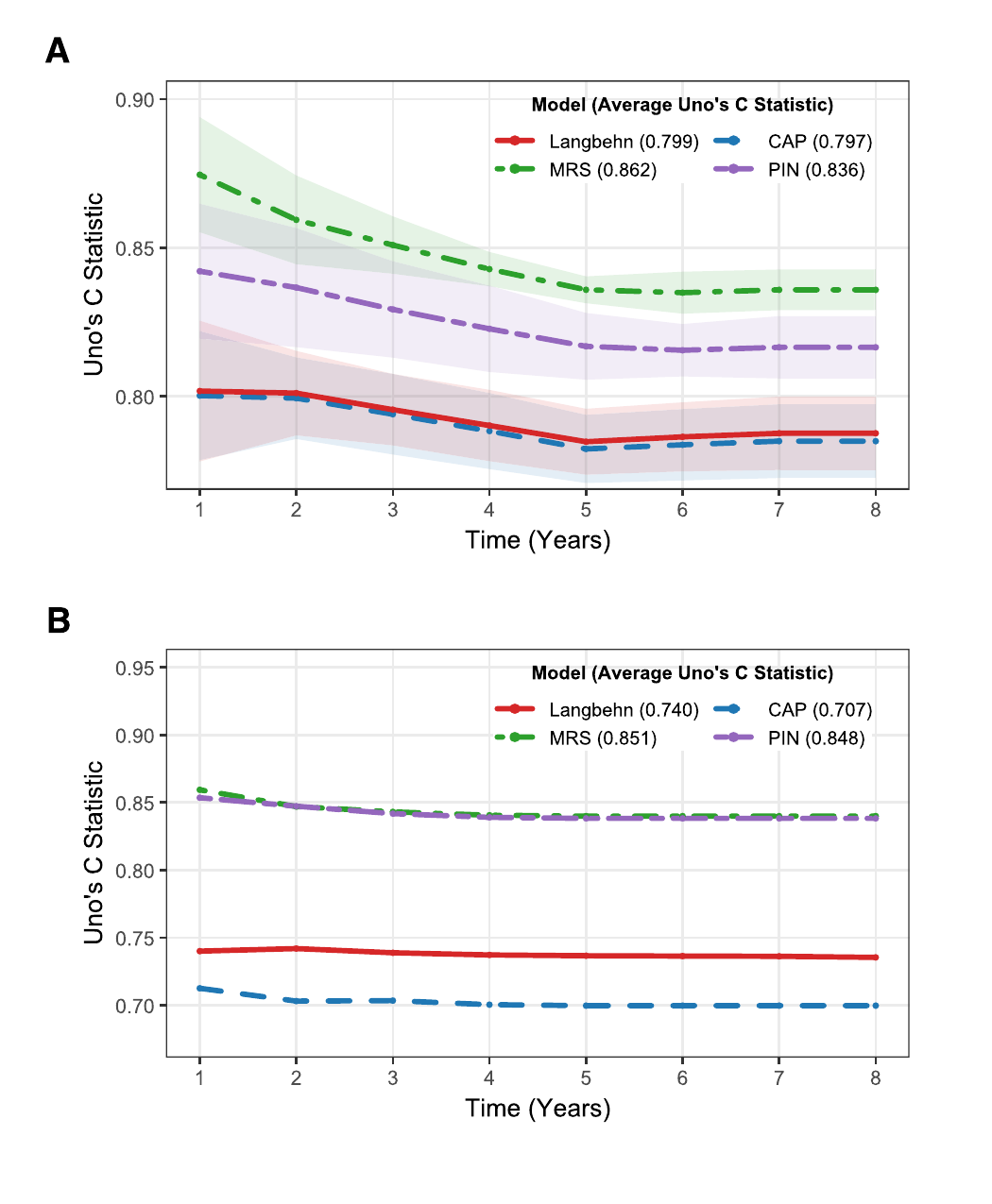}
    \caption{Time-specific Uno's C statistics by model and evaluation time for both published parameters (A) and updated parameters (B), with statistic given by mean from $5$-fold cross-validation for updated parameters. The average Uno's C statistic is a weighted average of the time-specific values. We considered four time-to-diagnosis models: CAG-Age Product (CAP), Langbehn, Multivariate Risk Score (MRS), and Prognostic Index Normed (PIN).}
    \label{fig:stackUnoC}
\end{figure}

All four models demonstrated moderate-to-strong risk stratification abilities overall (Figure~\ref{fig:stackUnoC}(B)), with global Uno's C statistics narrowly ranging from $0.80$ (CAP model) to $0.86$ (MRS model). Interestingly, the models' ordering was very consistent across all time points. The MRS model outperformed PIN, which outperformed the nearly-tied CAP and Langbehn models. We also saw that all models tended to perform worse over time, with each model's maximum Uno's C statistic observed at $\tau=1$ year from enrollment. This gradual decline in performance could be because the covariates were only taken at baseline (i.e., treated as time-independent).  For example, the PIN model does not update predictions as more recent SDMT scores are collected, even though those values would be more informative over time.

Over time, the MRS model consistently had the best predictive ability, producing average Uno's C statistics (across folds) between $0.84$ and $0.88$. Thus, this model has an estimated probability of $84\%$--$88\%$ of correctly ranking patients, i.e., of assigning lower log-risk scores to patients who were diagnosed later in time and higher log-risk scores to patients who were diagnosed sooner. The PIN model's performance was only slightly below that of MRS, producing average Uno's C statistics between $0.82$ and $0.84$. Altogether, these results strongly support the use of the MRS and PIN models in stratifying patients by risk of HD diagnosis. The CAP and Langbehn models performed well, performing just slightly below PIN.

We also evaluated the four models with the published parameters applied to the entire analytic sample (Figure~\ref{fig:stackUnoC}(A)). We expected there to be differences in performance between the updated and published parameter analyses, as the published parameters were obtained using different datasets. However, the differences in Uno's C between the two analyses were within $1\%$ to $6\%$ across all time points, suggesting that the updated parameters did not noticeably improve model performance.

\subsubsection{Kaplan-Meier Adjusted ROC Curves}

\begin{figure}
    \centering
    \includegraphics[width=0.8\linewidth]{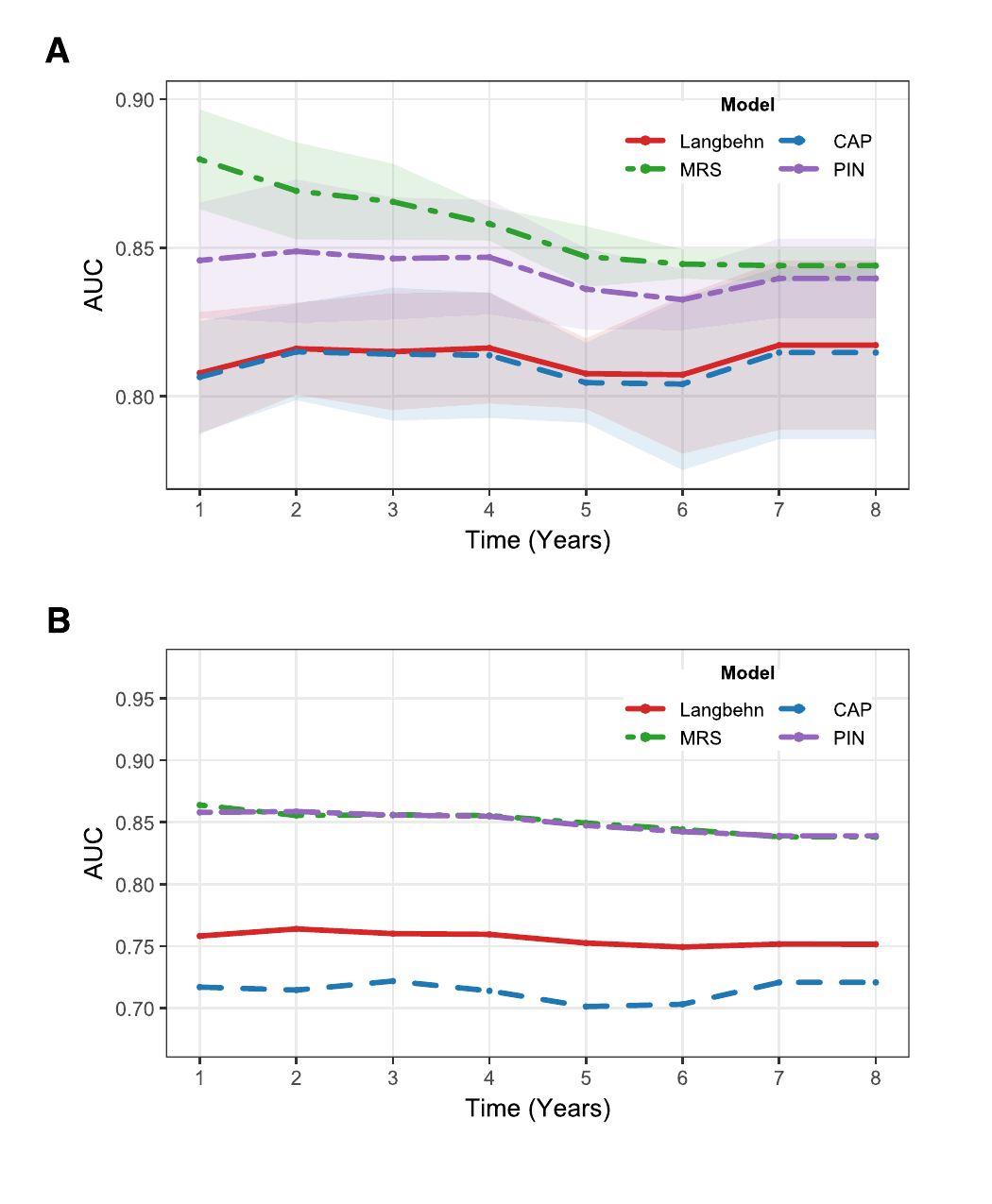}
    \caption{Time-specific area under the curve (AUC) of the receiver operating characteristic curve by model and evaluation time for both published parameters (A) and updated parameters (B). We considered four time-to-diagnosis models: Langbehn, CAG-Age Product (CAP), Multivariate Risk Score (MRS), and Prognostic Index Normed (PIN).}
    \label{fig:stackAUROC}
\end{figure}

As in the prior section, we calculated AUC values using $5$-fold cross validation to assess the models' performance when applied to data external to the training set. The models all produced AUC values similar to their Uno's C statistics, indicating similarly strong \textit{group-based} risk stratification (Figure~\ref{fig:stackAUROC}). The ordering of models based on their AUC values also matched the ordering based on Uno's C statistics, with MRS performing best, PIN performing second best, and Langbehn and CAP nearly tied. 

\subsection{Selecting an Optimal Log-Risk Score Threshold for Clinical Trial Sample Enrichment}
\label{sec:sample_size_est}

We leveraged Kaplan-Meier-adjusted ROC curves to inform sample enrichment for preventative clinical trials of HD. We used ROC curves to select an optimal log-risk score threshold, a necessary step toward sample enrichment, because they allowed us to consider the predictive accuracy of different possible thresholds (unlike Uno's C statistic). For simplicity, we used the published parameters to demonstrate this procedure. If, instead, parameters were to be updated, cross-validation would be needed here, as well. (Cross-validation could result in different thresholds and recommended sample sizes in each fold, which would need to be combined for implementation.)  

We exclude the Langbehn model from this portion of our analysis because it was difficult to implement and was tied with the CAP in terms of risk stratification. We considered four hypothetical trial durations ranging from $2$--$5$ years. For each combination of model and trial duration, we first calculated the \emph{optimal log-risk score threshold}. Using the Kaplan-Meier-adjusted ROC curve, we identified an optimal threshold $q^*$ that maximizes Youden's J statistic at time $t$, given by $J(q,t) = [\widehat{\textrm{TPR}}(q,t) + \{1 - \widehat{\textrm{FPR}}(q,t)\}- 1]$ \citep{Youden1950}. The enriched trial would enroll only patients whose estimated log-risk scores exceed $q^*$. 

Then, for the optimal threshold, we estimated the \emph{expected diagnosis rate in the untreated arm} (i.e., the proportion of patients who would be diagnosed with HD during the trial) among those whose who would be enrolled.
With this expected rate, we calculated the per-arm sample sizes required to detect treatment effects of $30\%$, $40\%$, and $50\%$ reductions in the rate of HD diagnosis, assuming $80\%$ power and a two-sided type-I error rate of $\alpha=0.05$. This sample enrichment strategy follows the same framework as Paulsen et al. (2019) \cite{Paulsen2019}.

Generally, the optimal thresholds for the CAP and MRS models were stable over time, while those for the PIN model varied more drastically (Table~\ref{tab:sample_size_estimates}). In addition, the expected diagnosis rates that correspond to these optimal thresholds tend to increase over time, which aligns with our expectation that more patients will be diagnosed in longer clinical trials. 
\begin{table}[ht]
\centering
\begin{tabular}{ccccccc}
  \toprule
\textbf{Time} & \textbf{Model} & \textbf{Threshold} & \textbf{Rate of Diagnosis} & $\mathbf{30\%}$ & $\mathbf{40\%}$ & $\mathbf{50\%}$ \\ 
  \midrule
  2 & CAP & \(370.20\) & \(0.32\) & \(343\) & \(186\) & \(114\) \\ 
  3 &     & \(370.20\) & \(0.37\) & \(272\) & \(148\) & \(92\) \\ 
  4 &     & \(370.20\) & \(0.40\) & \(245\) & \(134\) & \(83\) \\ 
  5 &     & \(368.46\) & \(0.41\) & \(236\) & \(129\) & \(80\) \\ 
\addlinespace
  2 & MRS & \(8.96\)   & \(0.36\) & \(289\) & \(157\) & \(97\) \\ 
  3 &     & \(8.96\)   & \(0.42\) & \(227\) & \(124\) & \(77\) \\ 
  4 &     & \(9.00\)   & \(0.46\) & \(200\) & \(110\) & \(68\) \\ 
  5 &     & \(8.73\)   & \(0.43\) & \(217\) & \(119\) & \(74\) \\ 
\addlinespace
  2 & PIN & \(0.41\)   & \(0.37\) & \(278\) & \(152\) & \(94\) \\ 
  3 &     & \(0.23\)   & \(0.39\) & \(257\) & \(140\) & \(87\) \\ 
  4 &     & \(0.27\)   & \(0.42\) & \(227\) & \(125\) & \(77\) \\ 
  5 &     & \(0.09\)   & \(0.40\) & \(249\) & \(136\) & \(84\) \\ 
\bottomrule
\end{tabular}
\caption[Summary of Sample Size Analyis]{Per-arm sample size estimates and optimal thresholds per model and trial duration. Effect sizes of 30\%, 40\%, and 50\% refer to relative reductions in diagnosis rate due to treatment. We considered three time-to-diagnosis models: CAG-Age Product (CAP), Multivariate Risk Score (MRS), and Prognostic Index Normed (PIN).} 
\label{tab:sample_size_estimates}
\end{table}
Although the optimal thresholds were not directly comparable across models due to their different scales, we compared the corresponding expected diagnosis rates. The optimal thresholds for the MRS model tended to yield the highest expected diagnosis rates, leading to the smallest required sample sizes among the three models at for trial durations of $3$--$5$ years. We focus on the MRS model, which had the strongest risk stratification ability (Section~\ref{sec:compare_risk_strat}). 

The optimal threshold for a $3$-year clinical trial was estimated to be $8.96$ for the MRS model. Therefore, we would aim to recruit patients whose estimated log-risk scores were $\geq8.96$. In doing so, we would expect $42\%$ of the untreated group to be diagnosed with HD during the trial. We estimated that $77$ patients per arm would be required to detect a $50\%$ treatment effect with $80\%$ power and $\alpha=0.05$. For smaller treatment effects of $40\%$ and $30\%$, we obtained per-arm sample size requirements of $124$ and $227$ patients, respectively. 

In prior work, Paulsen et al. (2019) \cite{Paulsen2019} estimated smaller per-arm sample sizes for each of these effect sizes: $67$ at $50\%$, $103$ at $40\%$, and $179$ at $30\%$. Although we used the same published parameter estimates for the MRS model, our analysis differs from that of Paulsen et al. (2019) \cite{Paulsen2019} in two key ways. First, we used data from ENROLL-HD, whereas Paulsen et al. (2019) \cite{Paulsen2019} used PREDICT-HD, the same dataset on which the CAP, MRS, and PIN models were originally developed to some extent. Second, our ROC analysis adjusted for censoring, which was crucial given that the ENROLL-HD data were $78\%$ censored, whereas the prior work did not. Compared to our censoring-adjusted, externally-validated calculations, the sample sizes estimates from Paulsen et al. (2019) \cite{Paulsen2019} would result in an underpowered clinical trial, thereby increasing the probability of failing to detect a truly effective HD treatment.

\section{Discussion}\label{discuss}

Multiple models have been proposed to predict the time to HD diagnosis, varying in their assumptions and inputs. To date, there have been few direct comparisons of these models, and those that exist have methodological limitations. Namely, they evaluated the models on the same data used to develop them (double-dipping) and used conventional performance metrics for comparison (failing to account for heavy censoring). We applied the Langbehn, CAP, MRS, and PIN models to data from the ENROLL-HD study, which was not used to develop any of the four models. We then compared each model's predictive performance using Uno's C statistic and Kaplan-Meier-based ROC analysis, which accounted for the high proportion of censored data. With these censoring-appropriate metrics, we also demonstrated how these models can be used in sample enrichment for a preventative clinical trial.

In comparing model performance via $5$-fold cross-validation (Section~\ref{sec:compare_risk_strat}), we observed a clear trade-off between risk stratification and parsimony. The MRS model demonstrated the strongest risk stratification capability but was the most complex (requiring the most covariates [eight]). Meanwhile, the PIN model, which required half as many covariates (four), performed only $3\%$ worse than MRS. The CAP and Langbehn models, using only two covariates, tie for third place, each performing about $4\%$ worse than PIN. However, based on our experiences, we recommend against using the Langbehn model: It is the most computationally complex and is no more parsimonious than the CAP model since both require the same covariates yet their performance is nearly identical. If parsimony is a priority, the CAP model is simpler to implement and interpret. If more clinical measurements are available, the MRS and PIN models outperform both. Crucially, our ordering matched that of Paulsen et al. (2019) \cite{Paulsen2019}, so our work strengthens their arguments.

Altogether, these observations allow for a straightforward recommendation in the context of risk stratification. Use the MRS model if DCL, the three Stroop tests, SDMT score, TMS, age, and CAG repeats are all available. Second, use the PIN model if SDMT score, TMS, age, and CAG repeats are available but DCL and/or the Stroop test scores are missing. Use the CAP model if only age and CAG repeats are available. 

Estimating the CAP, MRS, and PIN models was straightforward using the \texttt{survival} package in R \citep{Therneau2022}. Estimating the Langbehn model was more difficult because of its unique structure, and no off-the-shelf software existed. We implemented an MLE procedure for this model in R, which is available online at \url{www.github.com/sarahlotspeich/hd_ttd_modeling}. A grid search over possible values of one of the parameters ($\beta_2$ in \eqref{eqn:Langbehn_model}) was needed, since it would otherwise converge to its lower bound during optimization. 

Since we focus on sample enrichment for preventative clinical trials, where we recruit patients that are nearer to diagnosis than others, the models' ability to stratify potential patients by risk was paramount. However, one could also compare these models in terms of their calibration, which refers to the strength of agreement between a model's predicted survival probabilities and the observed rates of diagnosis. Park et al. (2021) \cite{Park2021} provide a thorough review of metrics for assessing a survival model's calibration. Calibration plots provide an intuitive, graphical assessment. The Brier score, which simultaneously quantifies both risk stratification and calibration, could also be used, though it would have to be adjusted for censoring (e.g., through IPCW) if applied to HD data.

When calculating the optimal thresholds and per-arm sample sizes for a preventative clinical trial, we applied each model's published parameter estimates to our entire analytic sample. Performing these calculations with the updated parameters via a cross-validation procedure, similar to Section~\ref{sec:compare_risk_strat}, would also be straightforward. 

Another avenue for future development is to build upon the best-performing model identified here. For example, 
all models tended to perform worse at later time points. This later-stage performance could potentially be improved by incorporating time-dependent covariates into the models (e.g., changing TMS and SDMT scores over time). Further, in preventative clinical trials, patients are recruited pre-diagnosis, which largely eliminates concerns about competing risks. However, trials targeting patients at later disease stages would need to consider incorporating a competing risk analysis (e.g., for censoring due to death from HD-related complications before receiving a formal clinical diagnosis). 

\section*{Acknowledgments}
This research was supported by the National Institute of Neurological Disorders and Stroke grant R01NS131225, and the National Science Foundation Graduate Research Fellowships Program (GRFP).\vspace*{-8pt}

\section*{Supporting Information}

Web Appendix A referenced in Section~\ref{sec:compare_risk_strat} is available online with  R code estimating the different models and evaluating their performance at \url{www.github.com/sarahlotspeich/hd_ttd_modeling}.

\bibliographystyle{unsrt}  
\bibliography{bibliography/biblio}

\label{lastpage}

\end{document}